\documentclass[pdflatex,sn-apa]{sn-jnl}

\usepackage{graphicx}%
\usepackage{multirow}%
\usepackage{amsmath,amssymb,amsfonts}%
\usepackage{amsthm}%
\usepackage{mathrsfs}%
\usepackage[title]{appendix}%
\usepackage{xcolor}%
\usepackage{textcomp}%
\usepackage{manyfoot}%
\usepackage{booktabs}%
\usepackage{algorithm}%
\usepackage{algorithmicx}%
\usepackage{algpseudocode}%
\usepackage{listings}%

\usepackage{array}
\usepackage{geometry}

\usepackage{forest}%

\usepackage{todonotes}

\theoremstyle{thmstyleone}%
\theoremstyle{thmstyletwo}%

\theoremstyle{thmstylethree}%

\begin{document}

\title[Article Title]{Automated Assessment in Mobile Programming Courses: Leveraging GitHub Classroom and Flutter for Enhanced Student Outcomes}


\author*[1]{\fnm{Pedro} \sur{Alves}}\email{pedro.alves@ulusofona.pt}

\author[1]{\fnm{Bruno} \sur{Pereira Cipriano}}\email{bcipriano@ulusofona.pt}

\affil*[1]{\orgname{Lusófona University}, \orgaddress{\street{Campo Grande, 376}, \city{Lisbon}, \postcode{1700-921}, \country{Portugal}}}

\abstract{The growing demand for skilled mobile developers has made mobile programming courses an essential component of computer science curricula. However, these courses face unique challenges due to the complexity of mobile development environments and the graphical, interactive nature of mobile applications. This paper explores the potential of using GitHub Classroom, combined with the Flutter framework, for the automated assessment of mobile programming assignments. By leveraging GitHub Actions for continuous integration and Flutter’s robust support for test automation, the proposed approach enables an auto-grading cost-effective solution. We evaluate the feasibility of integrating these tools through an experiment in a Mobile Programming course and present findings from a student survey that assesses their perceptions of the proposed evaluation model. The results are encouraging, showing that the approach is well-received by students.}




\maketitle

\section{Introduction}\label{sec_introduction}

The increasing prevalence of mobile applications has led to a growing demand for skilled mobile developers. Consequently, mobile programming courses have become a crucial component of computer science curricula. However, these courses present unique challenges due to the inherent complexity of mobile development, stemming not only from the intricate design and implementation of mobile applications \citep{francese2015using} but also from the need to set up a complex development environment (IDEs, SDKs, emulators, etc...) \citep{modesti2021script}. The difficulty of automating the build and testing processes further complicates the teaching and assessment of mobile programming. Traditional methods of code validation, such as output matching and unit tests, are often inadequate for mobile applications, which are predominantly graphical and interactive. For example, to test the navigation within the application, tests must be executed within a graphical emulator which typically take a long time to run and are difficult to automate. This complexity is magnified in advanced courses where students must develop applications that interface with external sources such as remote APIs, sensors, and local databases \citep{paiva2022automated}.

Automated assessment tools\footnote{Also known as auto-graders} (AATs) have been successfully implemented in many programming courses to increase student autonomy and motivation \citep{enstrom2011five} while reducing the teacher’s workload \citep{ihantola2010review}. These tools can validate the correctness of student submissions and, in some cases, also assess its code quality \citep{cipriano2022drop,heckman2018developing}. Even though most AATs were designed for simpler programming exercises \citep{cipriano2024bridging}, some are well-prepared for project-based assignments, involving multiple files and integration with Git \citep{cipriano2022drop}. However, the application of AATs in mobile programming courses remains limited due to the aforementioned complexities. The graphical nature of mobile applications and the intricate connections between the user interface (UI) and the underlying model pose significant challenges for automated testing. As a result, there is a notable gap in the research and practice of using AATs for mobile programming education.

In recent years, the advent of multi-platform mobile development frameworks such as React Native\footnote{https://reactnative.dev/} and Flutter\footnote{https://flutter.dev/} has provided new opportunities for overcoming these challenges. Flutter, in particular, offers robust support for test automation, including both unit, widget and integration (instrumented) tests that can easily be executed from the command line. In particular, widget tests enable UI testing without the need for an emulator. This capabilities suggests that it might be feasible to implement automated assessment in mobile programming courses using Flutter. Moreover, by leveraging GitHub Classroom to manage student projects and GitHub Actions to automate the execution of tests, educators can create a more streamlined and effective assessment process. This approach not only prepares students for modern development practices but also enhances their learning experience by enabling rapid feedback through features like hot-reload.

Given these developments, this study seeks to explore the potential of using Flutter and GitHub Classroom for automated assessment in mobile programming courses. Specifically, it addresses the following research questions:

\begin{itemize}
    \item RQ1: To what extent can Flutter, GitHub Classroom and GitHub Actions be
integrated to support the automated assessment of mobile programming assignments?
    \item RQ2: What are the students' perceptions of the proposed evaluation model?
\end{itemize}

This study makes the following contributions:

\begin{itemize}
    \item Presents a novel approach for assessing programming assignments focused on developing mobile applications
    \item Presents the results of a student survey evaluating the proposed assessment method
\end{itemize}

This paper is structured as follows. Section~\ref{sec_introduction} introduces the challenges of mobile programming education and the need for automated assessment tools. Section~\ref{sec_background} provides background on mobile development, testing frameworks, and the use of GitHub Classroom for managing and assessing programming assignments. Section~\ref{sec_related_work} reviews related work on automated assessment in mobile programming. Section~\ref{sec_experience} presents the "awesome quotes" exercise, outlining its academic context, design, test implementation, automation process, and the challenges faced. Section~\ref{sec_results} summarizes the results of the student assessment and survey, highlighting their perceptions of the model. Finally, Section~\ref{sec_conclusion} concludes with a discussion of the findings, the model's effectiveness, and potential improvements for future research.

\section{Background}\label{sec_background}

\subsection{Mobile development}\label{sec_mobile_development}

Currently, developing mobile applications involves targeting the two dominant operating systems (OSs): iOS and Android, which together account for over 99\% of the market share \citep{hayat2024}. When these OSs first emerged, the only viable model was native development, which required using the native SDK of each OS. For iOS, this involves developing the application in Objective-C or Swift, and the developer must have a computer running macOS since the SDK only runs on this operating system. For Android, this involves developing in Java or Kotlin, but the SDK runs on any operating system (Windows, macOS, or Linux). Given the significant differences between the SDKs, the only solution is to develop two completely separate applications, one for iOS and another for Android, with little opportunity for code reuse, resulting in increased costs. Consequently, some companies opt to initially develop only for one OS.

Meanwhile, three alternative models have emerged that allow development for both OSs from a single codebase: the so-called cross-platform models.

The simplest model is to develop the application in HTML5, which allows responsive screens to be designed using CSS3, adapting to different screen sizes to ensure good usability even on smaller devices. Additionally, HTML5 capabilities have evolved, providing limited access to sensors, geo-location, local storage, etc.\footnote{See https://whatwebcando.today/}. The main advantage of this model is its simplicity and ease of learning. However, it requires connectivity when the application is launched and, most importantly, does not allow the application to be installed from app stores, which users are accustomed to and expect.

Next, hybrid models emerged: models that allow the creation of installable applications for both OSs from a single codebase using an intermediate language. These hybrid models are divided into two types: hybrid-web and hybrid-native models.

Hybrid-web models, with the Ionic\footnote{https://ionicframework.com/} framework being the most popular representative, are essentially a "trick" because although they are installable applications, they are actually web applications embedded in a \textit{webview}, the browser component that renders pages. For this reason, they are developed in HTML and JavaScript, making them particularly appealing to web designers. The major disadvantage is that, not being truly native, they have limitations in accessing certain device functionalities and cannot guarantee the same fluidity and performance as native applications.

In response to these limitations, hybrid-native models emerged, with React Native and Flutter being the most popular representatives. These models use intermediate languages but produce binaries that run natively on their respective operating systems, as if they were native applications. They therefore offer equivalent fluidity and performance to native applications and nearly unlimited access to the device's native functionalities. Additionally, they provide a superior development experience thanks to hot-reload mechanisms, which eliminate the need to recompile/reinstall the application each time a code change is made. For these reasons, these models have gained popularity in recent years and are already adopted by major companies such as Meta, Google, Uber, Shopify and Alibaba. While React Native and Flutter have both established themselves as leading cross-platform frameworks in recent years, Flutter appears to be gaining momentum and outpacing React Native \citep{statista2024}.

\subsection{Automatic tests for mobile applications}

\subsubsection{Types of tests}

This section presents some types of tests which are relevant for mobile application development.

\textbf{Unit-tests} focus on testing a single function, method or class. They are usually used to test business logic.

\textbf{UI Component tests} (also referred to as \textit{widget tests}) are focused on checking if individual widgets are responding as expected (e.g., ensuring that adding text to a widget makes that text accessible in another widget).

\textbf{Integration tests} are focused on testing the interaction between components. In the specific case of mobile applications, the screens/activities/views can be seen as `components', although other components also may exist (e.g. database).

\subsubsection{Testing frameworks}
\label{sec:testing_frameworks}

\textbf{Unit testing frameworks} are usually built around the concept of \textit{assertions}. These frameworks provide various assertion methods that verify whether an expression (e.g., a variable's value or a function's return value) meets a specified condition. If the condition is not met, an error is reported. For example, when executing a test containing this assertion \texttt{assertTrue(backAccount.canWithdraw(10))} will issue an error if the \texttt{canWithdrawal()} function returns \texttt{false}.

JUnit is commonly used for unit testing of Android applications, while XCTest provides equivalent functionality for testing iOS applications.

\textbf{UI Testing frameworks}

UI Testing frameworks typically support UI Component tests and Integration tests. In this section, we we explore some UI Testing frameworks which should be considered when selecting a UI Testing framework for mobile applications.

\begin{description}
    \item[Espresso] is an automated testing framework specifically developed by Google for testing Android applications \citep{espresso-web-site}. It allows developers to write UI tests by providing a simple API. Tests can be written in either Java or Kotlin. Espresso allows interacting with widgets via code. See Listing \ref{lst:espresso_example} for an example of an Espresso UI test. Typically, these tests use a widget's ID to find it within a View, and then emulate actions on that widget, such as writing text on text fields, clicking on buttons, and so on.
    \item[XCUITest] is Apple's native framework for automating UI testing of iOS applications. It is tightly integrated with Xcode, providing a seamless environment for developers to write and execute tests specifically for iOS. XCUITest leverages Swift or Objective-C, the same languages used for iOS development, which allows developers to create tests that interact directly with the app's UI elements. Listing \ref{lst:xcuitest_example} shows an example test written in this framework.
    \item[Selendroid] is an open-source test automation framework designed for Android applications, offering a similar functionality to Selenium --- an open source test automation framework enabling developers and testers to automate browser interactions --- but tailored for mobile environments \citep{selendroid-web-site}. Tests for Selendroid can be scripted in multiple languages such as Java/Kotlin, C\#, Python, and Ruby \citep{risticdevelopment}.
    \item[Appium] is an open-source automated testing framework designed for mobile applications, providing a common API that allows developers to write tests for both Android and iOS platforms \citep{wang2019research}. Tests can be written in a variety of languages such as Java/Kotlin, C\#, Ruby and Python \citep{risticdevelopment}. One of the key strengths of Appium is its flexibility: it does not require any modifications to the source app and supports multiple languages as well as platforms. 
Appium has a client-server architecture: the server offers an Application Programming Interface (API) to execute commands on a mobile device and obtain the corresponding results; the client implements the actual tests by performing actions through the UI elements in the mobile app.
    \item[flutter\_test] is a flutter library\footnote{https://docs.flutter.dev/testing/overview} included into the Flutter framework which supports various types of tests (e.g. unit tests, widget tests and integration tests). With flutter\_test, developers can simulate user interactions, verify UI elements, and ensure the correctness of their app's functionality in different scenarios. The framework provides tools to check that widgets render as expected, that business logic behaves correctly, and that the app's performance remains optimal.
\end{description}

Table \ref{tab:comparative-table} presents a comparison of these frameworks across multiple criteria.

\lstset{ 
  language=Java, 
  basicstyle=\ttfamily\small, 
  keywordstyle=\color{blue}\bfseries, 
  stringstyle=\color{red}, 
  commentstyle=\color{gray}, 
  numbers=left, 
  numberstyle=\tiny\color{gray}, 
  stepnumber=1, 
  numbersep=5pt, 
  showspaces=false, 
  showstringspaces=false, 
  showtabs=false, 
  frame=single, 
  tabsize=2, 
  breaklines=true, 
  breakatwhitespace=false, 
  captionpos=b 
}

\newpage
\begin{lstlisting}[language=Java, caption={Espresso Example Test written in Kotlin. This test emulates writing a String on a text field and pressing a button. Finally, it checks if a specific String (e.g. ``Hello Steve'') is displayed in the View.}, label={lst:espresso_example}]
@Test
fun greeterSaysHello() {
   onView(withId(R.id.name_field)).perform(typeText("Steve"))
   onView(withId(R.id.greet_button)).perform(click())
   onView(withText("Hello Steve!")).check(matches(isDisplayed()))
}
\end{lstlisting}



%
\begin{lstlisting}[language=Swift, caption={XCUITest Example Test written in Swift. This test emulates writing a String on a text field and pressing a button. Finally, it checks if a specific String (e.g. ``Hello Steve'') is displayed in the view.}, label={lst:xcuitest_example}]
func testGreeterSaysHello() {
   let app = XCUIApplication()
   app.launch()

   let nameField = app.textFields["name_field"]
   nameField.tap()
   nameField.typeText("Steve")

   let greetButton = app.buttons["greet_button"]
   greetButton.tap()

   let greetingLabel = app.staticTexts["Hello Steve!"]
   XCTAssertTrue(greetingLabel.exists, "The greeting text doesn't exist.")
}
\end{lstlisting}

\begin{table}[h!]
\centering
\small
\begin{tabular}{|p{2cm}|p{2cm}|p{3cm}|p{2.5cm}|p{1.5cm}|}
\hline
\textbf{Framework} & \textbf{Platform} & \textbf{Test Types} & \textbf{UI Test \newline Execution Speed} & \textbf{Run Tests \newline Without \newline Emulator} \\ \hline
Espresso  & Android & UI, Unit, Integration & Fast & No \\ \hline
XCUITest  & iOS & UI, Integration & Fast & No \\ \hline
Selendroid & Android & UI, Integration & Moderate & No \\ \hline
Appium    & Android, iOS & UI, Integration & Moderate to Slow & No \\ \hline
flutter\_test   & Android, iOS & UI, Unit, Integration & Fast & Yes \\ \hline
\end{tabular}
\label{tab:comparative-table}
\caption{Comparison of Mobile Testing Frameworks}
\end{table}

\subsection{GitHub Classroom}

\begin{figure}
    \centering
    \includegraphics[width=0.7\linewidth]{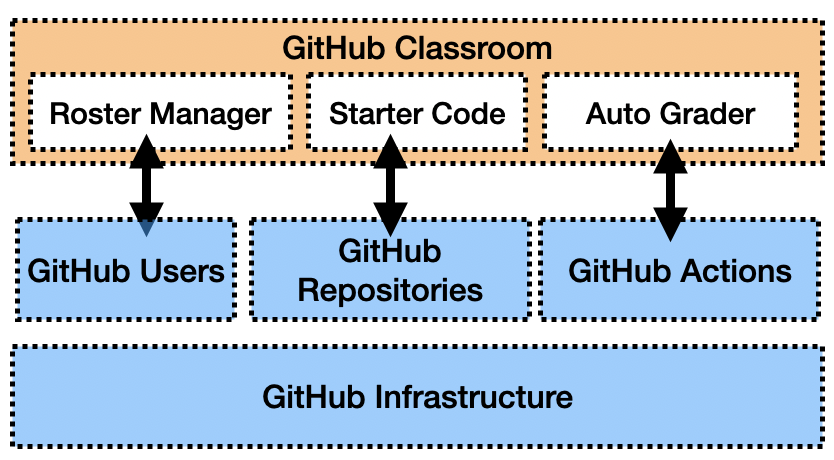}
    \caption{GitHub Classroom Architecture}
    \label{fig:GitHub-classroom-architecture}
\end{figure}

GitHub Classroom is a platform that integrates GitHub with educational workflows to facilitate the distribution, management, and assessment of programming assignments \citep{tu2022github,xu2023github}. As shown in Figure~\ref{fig:GitHub-classroom-architecture}, it is implemented as an additional layer on top of GitHub's existing features and infrastructure, enhancing them to manage the student roster, provide starter code for assignments, and include an auto-grader that runs tests on the code in the repositories. 

Classrooms must be associated with a GitHub organization, which is a shared account that allows multiple users to collaborate on repositories. There is no limit on the number of organizations one can establish. Within an organization, multiple classrooms can be created. For instance, an organization might be set up for a university department, with individual classrooms representing each course offered within that department. Each classroom has its own teaching assistants (TAs) and students.
Within each classroom, teachers create assignments and share them with the students through an invitation link.

We now describe the three main functionalities of GitHub Classroom.

\subsubsection{Roster management}

Since students utilize their own GitHub accounts, a mechanism is required to associate each account with the corresponding student ID. GitHub Classroom facilitates this association through a two-step process.

First, upon the creation of a classroom, the student roster can be imported either by connecting to the institution’s Learning Management System (LMS)\footnote{Supported platforms include Google Classroom, Canvas, Moodle, and Sakai}, or by uploading a file containing the students' IDs.

Subsequently, when students access an assignment within the classroom, they are prompted to select their ID from the previously imported student roster. It should be noted that this association is optional; students may choose to skip this step. If a student selects an ID, their GitHub account is linked to the corresponding student ID within that classroom. This association is only required once per classroom.

\subsubsection{Starter code}

When a student accepts an assignment\footnote{After clicking the invitation link, students are explicitly prompted to confirm their acceptance of the assignment}, GitHub Classroom automatically generates a repository within the organization, named after the student's GitHub account. Therefore it follows the \textit{one repository per student per assignment} distribution model, according to Glassey\citep{glassey2019adopting}. The repository is configured with appropriate permissions, granting access to both the student and the teachers. From this point, the student can push changes to the repository, and teachers can monitor the progress. Other students will not have access to these repositories unless the assignment is designated as public.

Typically, the newly created repository is not empty; it generally includes at least the assignment instructions (commonly provided in the \texttt{README.md} file) and files containing the testing code that will be used to evaluate the student's solution. In some cases, it also includes the project structure (e.g., folder hierarchy) as well as configuration/dependency management files (e.g., \texttt{pom.xml} for Maven projects or \texttt{pubspec.yaml} for Flutter projects). Additionally, the repository may contain source files with partial solutions. See \href{https://github.com/classroom-resources/autograding-example-java}{https://github.com/classroom-resources/autograding-example-java} for an example of such repository for a Java assignment, provided by GitHub.

When teachers create an assignment, they have the option to link it to a repository (referred to as the \textit{starter repository}), which serves as a template for all student repositories. This repository may have the files refered in the previous paragraph, and each student's repository is essentially a fork of this starter repository, providing several advantages. For instance, if modifications or corrections are needed in the \textit{starter repository} after students have accepted the assignment, the teacher can propagate these changes to all student repositories via the assignment dashboard. This action generates a Pull Request (PR) for each student repository, which the students must accept and merge to incorporate the updates into their own repositories.

\subsubsection{Autograder}
\label{sec:autograder}

This component is responsible for automating the execution of tests to evaluate and potentially grade the students' solutions. GitHub Classroom leverages GitHub Actions, which can be used as they are or adapted to better meet the specific needs of autograding.

We begin by providing an overview of GitHub Actions and then detail how they have been specifically adapted for autograding purposes.

\vspace{1ex} %
\noindent\textbf{GitHub Actions} 
\vspace{0.5ex}

GitHub Actions, officially released in November 2019, automates and orchestrates the execution of custom workflows directly in the repository. Using a YAML-based configuration (see Listing~\ref{lst:java-ci-maven} for an example), it defines workflows triggered by events such as code commits, pull requests, or scheduled intervals, enabling tasks like continuous integration (CI), continuous deployment (CD), testing, and code analysis\citep{tu2022github}. Workflows run in a virtual machine or container provided by the GitHub infrastructure.

\lstdefinelanguage{yaml}{
  keywords={name, on, push, branches, jobs, runs-on, steps, uses, with, run},
  keywordstyle=\color{blue}\bfseries,
  sensitive=false,
  comment=[l]{\#},
  commentstyle=\color{green}\ttfamily,
  stringstyle=\color{orange}\ttfamily,
  basicstyle=\ttfamily\small,
  morestring=[b]',
  morestring=[b]",
  literate =  *{:}{{\textcolor{red}{:}}}{1}%
               {,}{{\textcolor{red}{,}}}{1}%
               {[}{{\textcolor{red}{[}}}{1}%
               {]}{{\textcolor{red}{]}}}{1}%
               {>}{{\textcolor{red}{>}}}{1}, 
  breaklines=true,
  frame=single,
  captionpos=b,
  numbers=left,
  numberstyle=\tiny\color{gray},
}

\begin{lstlisting}[language=yaml, float=t, caption={GitHub Actions workflow for running the tests on a Maven/Java project. It is triggered by every push, runs in a container with the latest version of the Ubuntu operating system and it starts by cloning the repository, setting up the java environment and then executing maven tests.}, label={lst:java-ci-maven}]
name: Maven test

on:
  - push
  
jobs:
  test:

    runs-on: ubuntu-latest

    steps:
    - name: Clone repository into the runner environment
      uses: actions/checkout@v4
    - name: Set up JDK 17
      uses: actions/setup-java@v4
      with:
        java-version: '17'
        distribution: 'temurin'
        cache: maven
    - name: Run Maven Tests
      run: mvn clean test
\end{lstlisting}

Creating a YAML file to define workflows can be a complex task; however, there are numerous preset workflows available that can be readily used with minimal adjustments\footnote{See \href{https://github.com/marketplace?type=actions}{https://github.com/marketplace?type=actions}}.

\vspace{1ex} %
\noindent\textbf{Adapting GitHub Actions for autograding} 
\vspace{0.5ex}

GitHub Actions can be used for autograding directly by including a workflow YAML file\footnote{To be recognized by GitHub Classroom, this file must be named `classroom.yml`} in the \textit{starter repository}. Its biggest advantage is its flexibility: it can support virtually any programming language and integrate with any testing or code quality tool, provided that the tool can run within a virtual environment.

\begin{figure}
    \centering
    \includegraphics[width=0.9\linewidth]{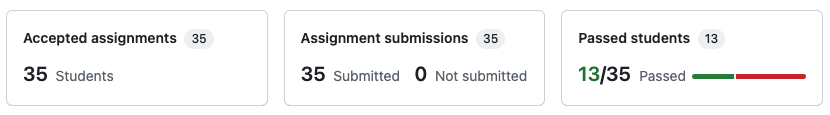}
    \caption{Partial view of the assignment dashboard}
    \label{fig:github-classroom-dashboard}
\end{figure}

GitHub Actions provide a binary outcome: \textbf{success}, when all jobs complete successfully, or \textbf{fail}, if any of the jobs exits with a non-zero code. GitHub Classroom processes this information to update the assignment dashboard, an aggregated view of grading results, displaying the number of students who accepted the assignment, submitted a solution, and passed the tests (see Figure~\ref{fig:github-classroom-dashboard}).

A major limitation of GitHub Actions is its lack of the granularity needed for effective autograding. For instance, in the workflow shown in Listing~\ref{lst:java-ci-maven}, consider a scenario with 20 tests where only one test fails. This outcome is significantly different from a scenario where all tests fail, yet both situations result in a workflow marked as \textbf{fail}. Although the exact number of passed tests can be determined by manually inspecting the workflow execution log, this process is cumbersome for both students and instructors, and it is not feasible to automate such detailed insights.

\begin{figure}
    \centering
    \includegraphics[width=0.5\linewidth]{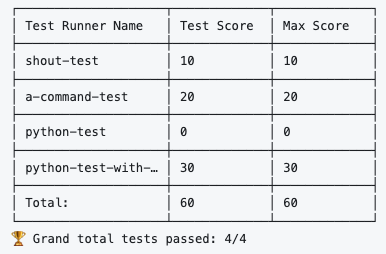}
    \caption{Example of the report produced by the Autograding Grading Reporter, combining the result of 4 different jobs.}
    \label{fig:autograding-grading-report}
\end{figure}

This limitation has been partially addressed by GitHub through the development of specialized GitHub Actions tailored for autograding purposes. For instance, the GitHub Action \href{https://github.com/classroom-resources/autograding-io-grader}{autograding-io-grader} allows for output matching tests and includes inputs such as a \textit{max-score} parameter, which represents the maximum points a student can earn for a given test. These can be combined with other test cases to provide a comprehensive grading solution. The key distinction of this action is its output format: instead of a binary result, it produces a JSON file containing detailed test results. Additionally, there is another GitHub Action, \href{https://github.com/classroom-resources/autograding-grading-reporter}{autograding-grading-reporter}, designed to process this JSON output and generate a detailed report for students, enhancing the feedback experience (see Figure~\ref{fig:autograding-grading-report}). However, only a limited number of these actions are available. As of August 2024, there are only two other actions specifically developed for autograding: autograding-command-grader\footnote{https://github.com/classroom-resources/autograding-command-grader} (for generic command line scripts) and autograding-python-grader\footnote{https://github.com/classroom-resources/autograding-python-grader} (for python projects).

Finally, the assignment dashboard provides the option to download a CSV file containing a report with URLs to all student repositories and their associated student IDs. However, it is noteworthy that this file does not include a column indicating the GitHub Action result ('success' or 'fail'). If autograding actions were used with a defined \textit{max-score}, the file will instead include 'points\_awarded' and 'points\_available' columns.

\section{Related work}\label{sec_related_work}

Even though the demand for mobile programming specialists has been growing, reflected in computer science curricula, teaching mobile application development remains challenging due to the graphical nature of these applications and their advanced functionalities, such as remote API access and interaction with sensors.

One area that is underrepresented in the literature is the use of automated assessment tools in these courses. Automating the testing of mobile applications is inherently difficult due to their complexity, which often results in evaluations being conducted manually \citep{sung2014mobile}.

A recent survey on AATs \citep{paiva2022automated} found only 4 solutions for mobile development. All solutions are specific to Android devices.

Bruzual \citep{bruzual2020automated} proposes an Android grader that performs the actual assessment of the online exercises. The grader is a Docker container that includes all software needed to execute and evaluate student submissions. Assessment is carried out by running exercise-specific unit tests on the Android app submitted in binary format, namely, as an Android application package (APK). 
Android exercises are assessed through Appium, an open-source test automation framework widely used for
testing native mobile apps, including those for Android (see Section~\ref{sec:testing_frameworks}). 
Directly testing the app binary is advantageous because it decouples testing from the underlying code. However, interacting with a precompiled application through screen events restricts the range of possible tests, making it somewhat comparable to output-matching tests. Furthermore, this grader is specific to Android.

Madeja \citep{madeja2017automatic} proposes a testing environment specific for Android applications, comprised of a static testing step and test pyramid step. The static testing involved validating that the UI used the correct identifiers (through XML parsing) followed by manual teacher validation of screenshots sent by the students. The test pyramid step is called that way because it executes: unit tests (JUnit + Mockito), integration tests (Roboletric) and UI tests (Espresso). The first two don't require an emulator but the authors found the Roboletric tests to be unreliable due to conflicts between different Android APIs. Also, the fact that database access was done using a singleton created problems with test creation since the database wasn't easily cleaned up before tests execution.
Also, they were not able to test a critical part of the project due to limitations of Roboletric.
The tests with Espresso not only required executing in the emulator but also had some architectural limitations. For example, it was necessary to manually delete the database in some cases.

Our approach enables true integration testing as the tests run in the same VM as the application. For instance, it allows inspecting the model and injecting test-specific dependencies (e.g., mocks). Additionally, it integrates directly with Git, providing more immediate feedback through GitHub Actions. Finally, it supports testing for both Android and iOS.

\section{The ``awesome quotes'' experience}\label{sec_experience}

In this section, we present our approach to automatically assess a mobile programming exercise, which we called ``awesome quotes''.

\subsection{Academic context}

This experiment was conducted as part of a Mobile Computing course during the second semester of the third year in the Computer Engineering bachelor's program at the university, in the academic year 2023/24. The course had 76 enrolled students, but only 63 participated in the evaluation process. The curriculum of this course focuses on the distinctive features and limitations of mobile computing when compared to traditional computing, covering aspects such as geo-location, sensors, autonomy, usability, connectivity, and security. Additionally, students explore and compare the four mobile development models referred in \ref{sec_mobile_development}. Laboratory sessions are dedicated to practicing Flutter development, with pre-recorded instructional videos provided before the labs. Students work on a group project in pairs with two deliverables, implemented throughout the semester, with most lab classes dedicated to advancing this project. Additionally, students complete several individual exercises designed to be finished within one or two lab sessions, which serve to reinforce certain topics that will be useful for the project.

It is important to note that prior to this course, students had experience with two other AATs: Drop Project \citep{cipriano2022drop}, an open-source tool for Kotlin/Java programming assignments, and Pandora, an in-house tool for C programming assignments. Drop Project supports submissions via both upload and git, whereas Pandora allows only upload-based submissions. In both tools, a set of teacher-defined tests is executed, but students have access only to the test results, not the test code itself.

\subsection{Switch from Kotlin to Flutter}

In previous iterations of the course, the primary project involved developing a native Android application using Kotlin, with submissions managed via GitHub Classroom. Native development in Android posed several challenges: the setup process was complicated, and some students’ computers, as well as the lab computers, were not powerful enough to provide a good programming experience. Additionally, the lack of hot-reload made the experimentation process very slow, and there were no automated tests. Finally, students with iPhone could not test the application except through the emulator. For these reasons, we decided to switch the project to Flutter in the current year (2023/24). Also, notice that GitHub Classroom was only utilized to facilitate repository access for the professors, without leveraging its validation features or providing a project skeleton.

\subsection{Description of the exercise}

\begin{figure}
    \centering
    \includegraphics[width=0.8\linewidth]{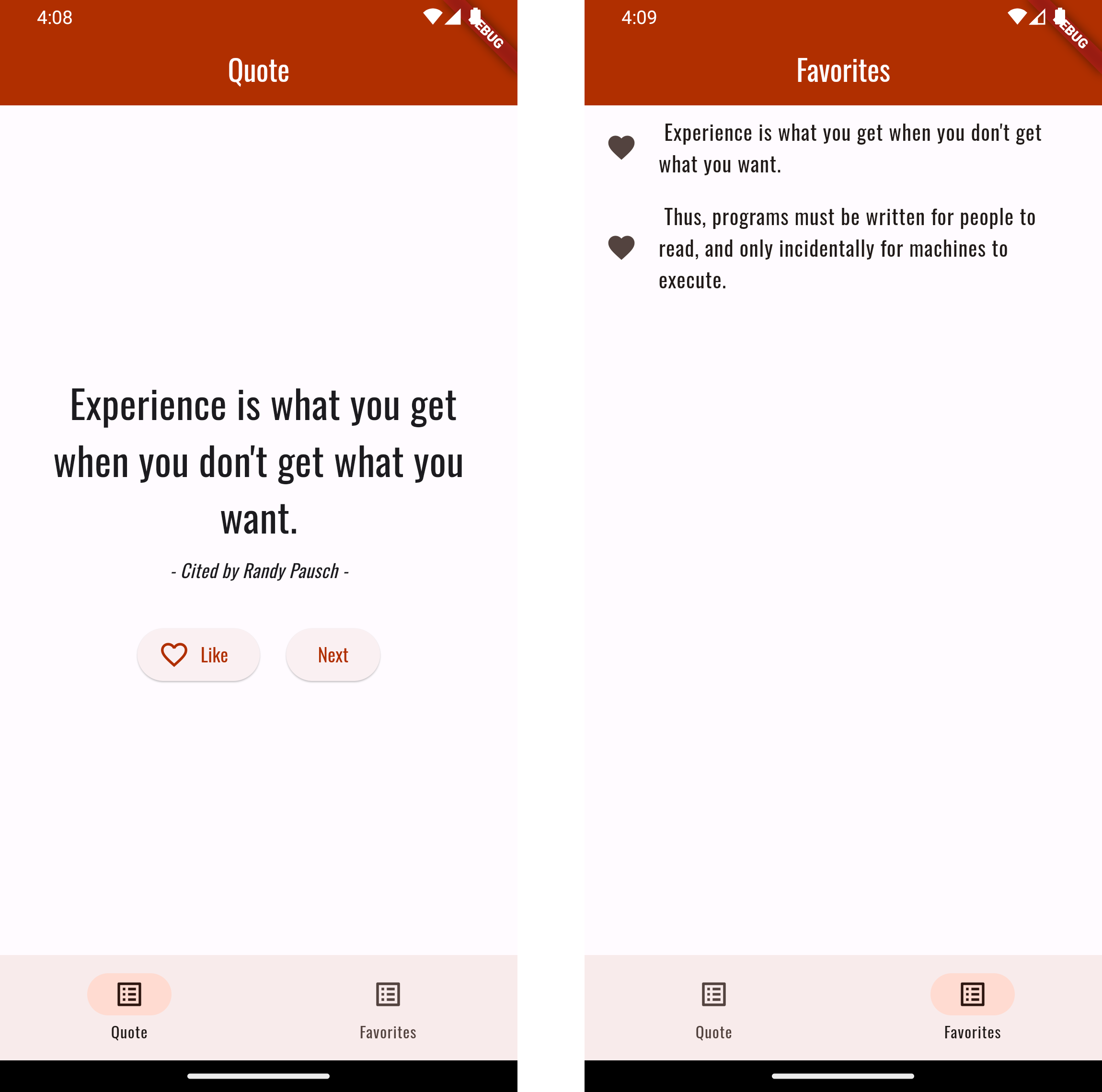}
    \caption{The ``awesome quotes'' application that the students were expected to implement. The screen on the left shows a random quote and the possibility to like it or generate another one. The likes quotes show up in the favorites screen, shown on the right.}
    \label{fig:awesome_quotes}
\end{figure}

In addition to switching to Flutter, we wanted to experiment with automated assessment. Rather than implementing this fully for the main project, we decided it would be wiser to start with a small exercise - the “awesome quotes” assignment. This optional exercise, offering a small bonus of up to 0.5 points on the final practical grade (out of 20), required students to develop a Flutter application that randomly displays inspiring quotes from a pre-loaded library. The application included two buttons for navigating quotes and marking favorites, and a secondary screen to list the favorite quotes (see Figure~\ref{fig:awesome_quotes}). Students had two weeks to complete this exercise.

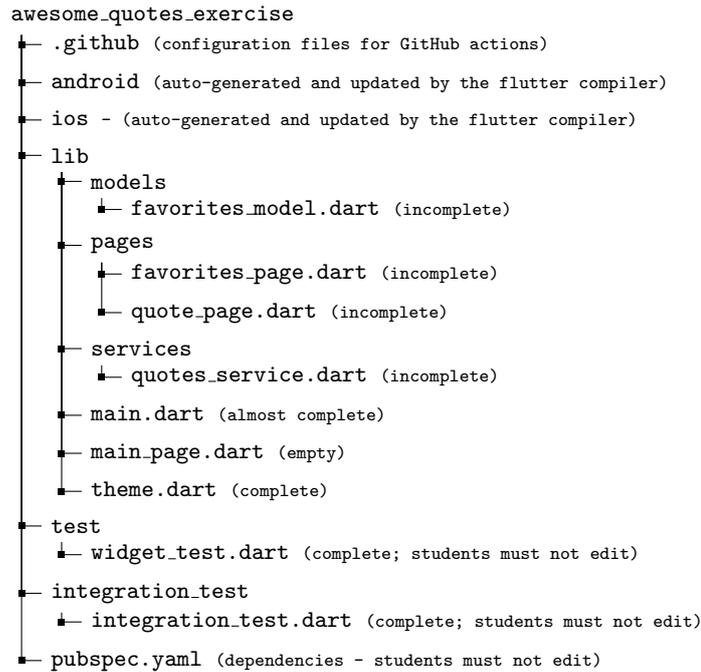
\begin{figure}[t]
    \centering
    \begin{forest}
    for tree={
        font=\small\ttfamily,
        grow'=0,
        child anchor=west,
        parent anchor=south,
        anchor=west,
        calign=first,
        edge path={
          \noexpand\path [draw, \forestoption{edge}]
          (!u.south west) +(7.5pt,0) |- node[fill,inner sep=1.25pt] {} (.child anchor)\forestoption{edge label};
        },
        before typesetting nodes={
          if n=1
            {insert before={[,phantom]}}
            {}
        },
        fit=band,
        before computing xy={l=15pt},
        l sep=0pt,
        s sep=0pt
      }
    [awesome\_quotes\_exercise
      [.github \footnotesize (configuration files for GitHub actions)]
      [android \footnotesize (auto-generated and updated by the flutter compiler)]
      [ios \footnotesize - (auto-generated and updated by the flutter compiler)]
      [lib
        [models
          [favorites\_model.dart \footnotesize (incomplete)]
        ]
        [pages
          [favorites\_page.dart \footnotesize (incomplete)]
          [quote\_page.dart \footnotesize (incomplete)]
        ]
        [services
          [quotes\_service.dart \footnotesize (incomplete)]
        ]
        [main.dart \footnotesize (almost complete)]
        [main\_page.dart \footnotesize (empty)]
        [theme.dart \footnotesize (complete)]
      ]
      [test
        [widget\_test.dart  \footnotesize (complete; students must not edit)]
      ]
      [integration\_test 
        [integration\_test.dart \footnotesize (complete; students must not edit)]
      ]
      [pubspec.yaml \footnotesize (dependencies - students must not edit)]
    ]
    \end{forest}
    \caption{The ``Awesome Quotes'' initial project structure. Most of the source code was incomplete, serving only as placeholders for the students to fill in. Test files and pubspec.yaml were complete and should not be edited by the student.}
    \label{fig:awesome_quotes_folder_tree}
\end{figure}

The ``awesome quotes'' exercise was distributed through GitHub Classroom. Students accepted the assignment via a provided link, which created a fork of the main repository in their GitHub accounts. They were instructed to clone the repository, open it in Android Studio, and implement the required functionality. The forked repository contained a skeleton of the recommended project structure (depicted in Figure~\ref{fig:awesome_quotes_folder_tree}) adhering to Flutter best practices, including several incomplete classes that students needed to finish. The repository also included a \texttt{pubspec.yml} file with required dependencies and several auto-generated files typical of a new Flutter project. Crucially, the repository featured a set of widget and integration tests that students were not allowed to modify but could run locally to verify their work. The exercise was deemed complete when all tests passed successfully. The full statement is publicly available in \href{https://github.com/palves-ulht/awesome_quotes_exercise}{https://github.com/palves-ulht/awesome\_quotes\_exercise}.

\subsection{Tests implementation}
\label{sec:tests-implementation}

Since the exercise had almost no business logic but rather it focused on the UI and in its interaction with the model, we decided not to implement unit tests, implementing only widget and integration tests.

One interesting fact is that the widget tests and integration tests were almost the same, with only a single line differentiating them, as shown in Listing~\ref{lst:integration-test}.

\begin{lstlisting}[language=C,basicstyle=\small\ttfamily,commentstyle=\color{gray}\ttfamily,label={lst:integration-test},float=t,caption={Example of a widget/integration test in Flutter. The only difference is the first line. Widgets tests execute without an emulator.}]
void main() {
  // this line only exists on integration tests
  IntegrationTestWidgetsFlutterBinding.ensureInitialized();

  // the rest of the code is the same for widget and
  // integration tests
  ...
  Text quoteText = tester.widget(find.byKey(kQuoteTextKey));
  String? quote = quoteText.data;
  expect(quote, isNotNull);

  await tester.tap(find.byKey(kNextButtonKey));
  ...
}  
\end{lstlisting}

We decided to have both tests because each one has unique advantages. The widget tests executed very fast, without needing an emulator. However, when they failed, it was easier to debug using integration tests since the programmer receives visual feedback during the tests, since they run in the emulator. For example, if the test fails because it doesn't find a certain widget in the page, it is much easier to see how the page is rendered in the emulator to understand the issue.

The students were encouraged to run both types of tests locally; however, GitHub Actions was configured to execute only the widget tests to conserve CPU cycles and avoid exceeding GitHub's usage limits.

Within each test file, we created 4 test functions, corresponding to 4 scenarios:

\begin{enumerate}
    \item Shows quote page, hit next and get a new quote. This was done using the real quote service, effectively getting a random quote each
    time the test ran and verifying if hitting next would get a different quote from the current one.
    \item Shows quote page. This was done using a fake quote service that always returned the same quote. The goal of this test was to check if the graphical elements existed and contained the correct information.
    \item Navigate to favorites and back to quote. This test checked if the navigation through the bottom bar was well implemented.
    \item Mark as favorite and navigate to favorites. This test checked the interaction between the UI and the model: tapping the 'like' button should add the current quote to the list of favorites (stored in the \texttt{FavoritesModel} class), and navigating to the 'favorites' page should show the update list of favorites.
\end{enumerate}

Note that some scenarios used the real quotes service while others used a fake quotes service. This was easily implemented because the students were instructed to follow a 'dependency injection' (DI) pattern \citep{fowler2004inversion}. Specifically, this means that widgets should receive a \texttt{QuotesService} object instead of instantiating/obtaining it themselves. This approach allowed the tests to inject different \texttt{QuotesService} objects as needed. Several libraries facilitate DI implementation; in this case, students used the Provider library\footnote{https://pub.dev/packages/provider}.

To properly test the widgets, students were required to use specific identifiers for each widget, as outlined in the exercise statement.

You can find the complete tests implementation in the exercise repository, publicly available at the following link: \href{https://github.com/palves-ulht/awesome\_quotes\_exercise}{GitHub Repository}\footnote{This is a copy of the original repository, with the statement translated to English}.


\subsection{Automating Tests}

The experiment incorporated GitHub Actions (configured in the \texttt{.github} folder, see Figure~\ref{fig:awesome_quotes_folder_tree}) to automate test execution with each push to the repository. This setup enabled continuous integration by running tests on GitHub’s servers, providing immediate feedback to students and simplifying the evaluation process for instructors. This automation ensured that professors could easily verify student submissions without manually cloning repositories and executing tests locally. Furthermore, students had to pass a code validation step using Dart’s static analyzer (\texttt{dart analyze}). Although these students had previous experience with AATs in earlier courses such as CS1 and CS2, this was their first exposure to using GitHub Classroom's integrated assessment capabilities.

See Listing~\ref{lst:awesome-quotes-github-action} for the content of the workflow file. Notice the three run jobs: installs dependencies, analyze the Dart code, and run Flutter tests. If any of these jobs fails, the workflow is marked as failed, leading to a negative assessment of the student’s submission. As already described in Section~\ref{sec:tests-implementation}, the 'run tests' job only executed widget tests to optimize for performance and reduce costs.

However, we also successfully experimented automating the integration tests running on a iOS Simulator, using a container running MacOS and \textit{simulator-action}\footnote{https://github.com/futureware-tech/simulator-action}, a GitHub action that helps you start an iOS Simulator inside the workflow you are running.

\begin{lstlisting}[language=yaml, caption={GitHub Actions workflow for the ``awesome quotes'' exercise. It is triggered by every push, runs in a container with the latest version of the Ubuntu operating system and it starts by cloning the repository, setting up the flutter environment, getting all the flutter dependencies ('pub get'), executing ´dart analyze' (a linter for dart) and then running flutter tests.}, label={lst:awesome-quotes-github-action}]
name: Run Tests

on: [push]

jobs:
  build:
    runs-on: ubuntu-latest

    steps:
      - uses: actions/checkout@v3
      - uses: subosito/flutter-action@v2
        with:
          flutter-version: '3.19.3'

      - run: flutter pub get
      - run: dart analyze
      - run: flutter test
\end{lstlisting}

\subsection{Challenges and limitations}
\label{sec:limitations}

Our experiment revealed some limitations of GitHub Classroom, particularly concerning execution limits (quotas) and the adequacy of provided testing information for grading assignments. We now describe both limitations.

\subsubsection{Execution quotas}
\label{sec:execution_quotas}

During the experiment (2024's second semester), GitHub Actions for private repositories were limited to 2000 minutes per month on the free plan. However, since academic accounts could be upgraded to the team plan, we were allocated 3000 minutes per month. We were concerned that students might abuse the system by making dozens of submissions, as they had done with other AATs, potentially exhausting the quota. However, this did not happen, likely because students had access to the test code and could run and debug the tests locally. On average, each student made 4.8 submissions, with each submission taking an average of 1 minute and 15 seconds. Since only 35 students accepted the assignment, the total execution time was 210 minutes, well below the limit.

However, if we had opted to run the integration tests using the iPhone simulator, the situation would have been different. MacOS runners consume 10 times more minutes than Linux\footnote{https://shorturl.at/3hk59}, effectively reducing the limit to 300 minutes per month. In our experiment, a single execution of the integration tests took 6 minutes and 34 seconds. Assuming the same average number of submissions, the total execution time would be approximately 1,103 minutes, far exceeding the limit.

\subsubsection{Testing reports from the teacher perspective}

As already mentioned in Section~\ref{sec:autograder}, testing reports collected by GitHub classroom only show whether the tests passed or failed, without providing details on the number or specific tests that failed. While this is not a major issue for students, who can run the tests locally to see the exact number of failures, it is very time-consuming for teachers to manually inspect each submission’s workflow log to determine the number of failed tests. We found this to be a significant limitation, since grading the ``awesome-quotes'' exercise was based on the number of tests passed. To address this, a custom GitHub Action could be developed to support autograding by converting the Flutter test output into a JSON format compatible with the autograding-reporter tool. However, due to time constraints and the lack of clear documentation, we were unable to develop such an action. 

Moreover, since this information (pass or fail) is not included in the CSV file available for download with aggregate results, grading had to be done manually for each submission.

\subsection{In summary}

The “awesome quotes” exercise, part of a Mobile Computing course, involved students developing a Flutter application, with GitHub Classroom used for repository distribution and GitHub Actions for continuous integration and automated testing. The setup enabled automated evaluation through widget tests configured to run each time students submitted code, ensuring immediate feedback. The experiment addressed \textbf{RQ1}, highlighting benefits such as simplified grading and reduced manual intervention, while also identifying challenges like execution time limits and limited reporting detail from GitHub Actions, which complicated grading based on test results.

\section{Results}\label{sec_results}

We now describe the results of the assessment through a student survey, highlighting their perceptions of this evaluation model.

35 students accepted the assignment. Of those, 14 passed all the 4 tests, with an additional 3 passing some tests.

\subsection{Survey}

To answer \textbf{RQ2}, we conducted an anonymous survey to gauge the students' perceptions of the exercise, after the exercise deadline. The survey focused on their overall experience, including motivation, clarity, and the usefulness of provided resources, as well as their preferences regarding automated versus manual assessment. A total of 27 students participated in the survey.

The survey began with a multiple-choice question to determine whether the students had participated in the exercise, followed by an open-ended question allowing those who did not participate to explain their reasons. The majority of the students accepted and attempted to solve the exercise. However, 4 students did not click the link to accept the exercise, and another 4 students accepted the exercise but did not attempt to solve it. Some reasons provided were related to initial configuration issues: ``\textit{I had errors that prevented the application from compiling, making it difficult to understand the working basis to achieve the given objective}'' and ``\textit{I couldn’t initialize the project in Android Studio. When I cloned it, it didn’t create a Flutter project but a regular project instead.}'' We suspect these students may not have attended the lab classes where the configuration of these projects was practiced.

All subsequent questions were directed at students who had accepted and attempted the exercise, as these are the primary focus of the survey. The students who participated in the exercise were asked 10 quantitative questions using a standard 5-point Likert scale (ranging from strongly disagree to strongly agree), followed by two qualitative questions where they could provide open-ended comments. The quantitative questions are described in Table~\ref{tab:survey_quantitative_questions}.

\begin{table}[]
\resizebox{\textwidth}{!}{%
\begin{tabular}{|l|l|}
\hline
Q. \# & Question                                                                                                                                                       \\ \hline
Q1    & I found the “awesome quotes” application interesting and motivating as an exercise.                                                                            \\ \hline
Q2    & \begin{tabular}[c]{@{}l@{}}I found the instructions clear and with enough information to solve the exercise \\ (including the referenced videos).\end{tabular} \\ \hline
Q3    & Having access to the test code (widget and integration) was useful for solving the exercise.                                                                   \\ \hline
Q4    & The code provided when creating the repository was sufficient to solve the exercise.                                                                           \\ \hline
Q5    & \begin{tabular}[c]{@{}l@{}}The submission model through GitHub, with automatic test execution \\ on the GitHub server, was clear to me.\end{tabular}           \\ \hline
Q6 &
  \begin{tabular}[c]{@{}l@{}}I prefer this exercise model (using GitHub Classroom with automatic test execution) over\\  the model where I only submit to GitHub and it is manually evaluated by the teacher \\ after the deadline (as used in the project).\end{tabular} \\ \hline
Q7    & Being able to run the tests locally makes the exercise more motivating.                                                                                        \\ \hline
Q8    & I find this model suitable for small exercises (1 or 2 screens).                                                                                               \\ \hline
Q9    & I find this model suitable for the project.                                                                                                                    \\ \hline
Q10 &
  \begin{tabular}[c]{@{}l@{}}For mobile computing exercises, I find this model more suitable than the ones \\ I used in programming courses (Drop Project, Pandora).\end{tabular} \\ \hline
\end{tabular}%
}
\caption{Survey quantitative questions}
\label{tab:survey_quantitative_questions}
\end{table}

\begin{figure}
    \centering
    \includegraphics[width=1.0\linewidth]{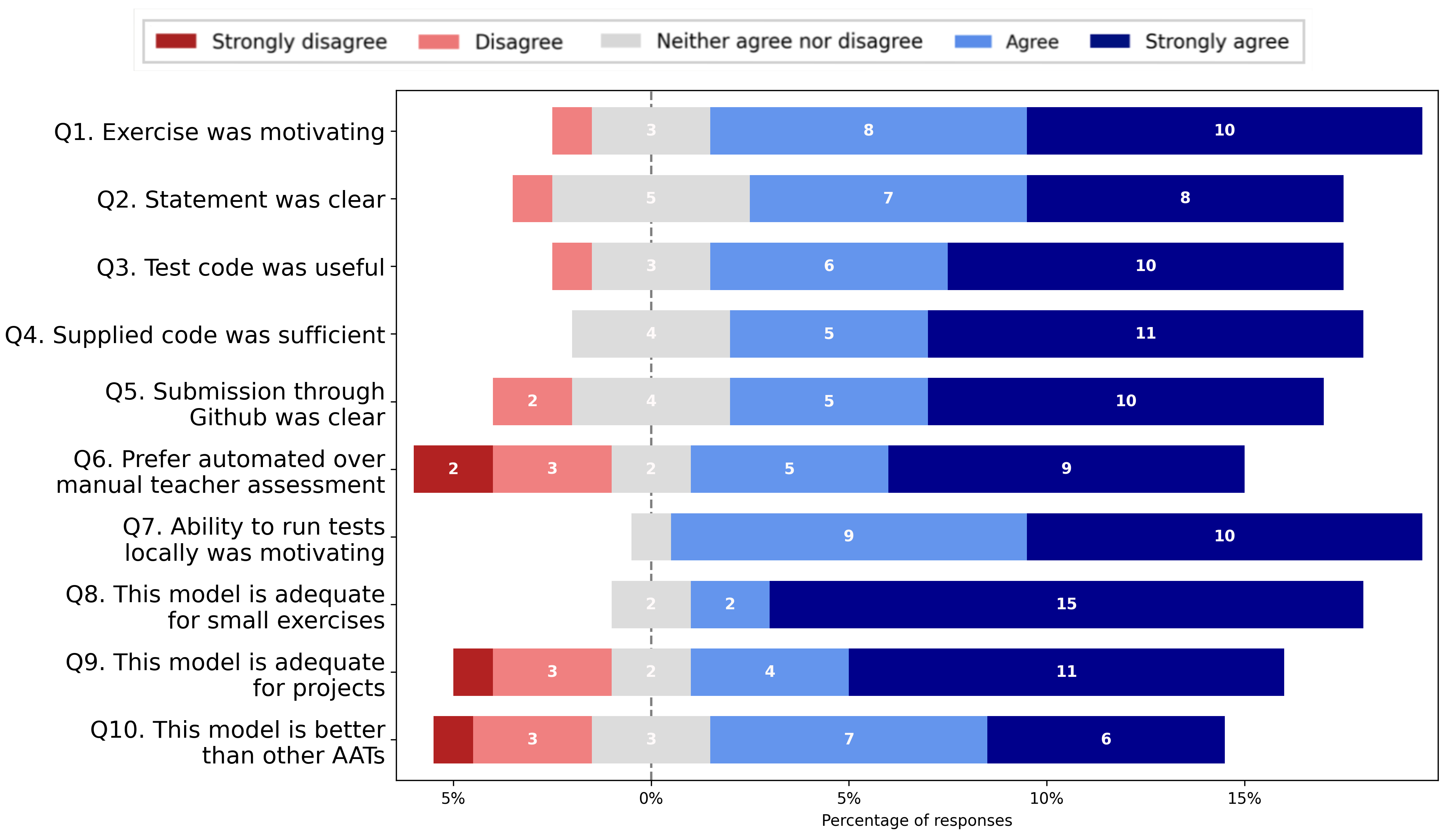}
    \caption{Results of an anonymous survey to the students, after the exercise deadline. The survey included 10 questions using a Likert scale.}
    \label{fig:survey_chart}
\end{figure}

The results of the quantitative questions are illustrated in Figure~\ref{fig:survey_chart}.

The first of these questions sought to determine whether the exercise was motivating for the students, specifically if the "awesome quotes" application was an interesting project. The majority of the students agreed or strongly agreed that the application was motivating.

Additionally, most students agreed or strongly agreed with the statements ``I found the statement clear and with enough information to solve the exercise (including the referenced videos),''(Q2) and ``Having access to the test code (widget and integration) was helpful in solving the exercise.''(Q3). These responses indicate that the format of the exercise, including the provided instructions and tests, was adequate.

As previously mentioned, the forked repository included some incomplete classes. There is a delicate balance between providing too little and too much code; the former can lead to student frustration, while the latter can reduce learning efficacy. To determine if this balance was appropriate, we asked students if they agreed with the statement, ``The code provided when creating the repository was sufficient to solve the exercise.''(Q4). Most students agreed or strongly agreed (80\%), with only a small fraction remaining neutral.

Given that students were not accustomed to validating their solutions using GitHub Classroom actions, it was important to assess if the process was clear. Most students agreed or strongly agreed with the statement, ``The submission model through GitHub, with automatic test execution on the GitHub server, was clear to me.''(Q5). Although we did not specifically ask about the quality of the test feedback, it is noteworthy that the absence of a detailed feedback report (as described in Section~\ref{sec:autograder}) did not appear to negatively impact the submission experience. This may be a result of students having access to the test code and the ability to run and debug these tests locally.

The next question (Q6) addressed the impact of automated assessment compared to traditional manual assessment by the teacher. Some of the course assignments had previously been submitted via GitHub but were evaluated manually by the TAs. Since this was the first assignment to be automatically graded, we inquired whether students preferred the automated model. The majority of students agreed or strongly agreed that the automated assessment was better; however, 5 students (24\%) disagreed or strongly disagreed with this preference.

The question ``Being able to run the tests locally makes the exercise more motivating'' (Q7), was inspired by the functionality of AATs used by these students in other courses. These AATs typically conceal the content of the tests and do not allow students to run them locally, which has led to some complaints among students, as informally noted in conversations. Our goal was to assess the actual impact of this feature on student motivation. The results confirmed the complaints, with all but one student (who was neutral) finding the ability to run tests locally to be more motivating.

Questions Q8 and Q9 aimed to explore the relationship between this assessment model and the nature of the assignments. Assignments can be exercise-based—consisting of small coding tasks with predetermined answers, typically completed in under an hour—or project-based, involving the development of a full application, which usually takes several days or weeks to finish. Some assessment tools are better suited for exercise-based assignments, while others are more appropriate for project-based tasks \citep{cipriano2024bridging}. Our goal was to determine where this model aligns most effectively. All the students except two (who remained neutral) found the model suitable for small exercises. While the majority (71\%) also deemed the model appropriate for projects, 4 students (19\%) disagreed, and 2 remained neutral. Overall, students considered the model appropriate for both types of assignments, though there was a stronger consensus on its suitability for small exercises.

Finally, we asked whether ``This model is better than other AATs'' (Q10) to benchmark this approach against other AATs that the students had used in other courses. Most students (65\%) considered this model better, while only 20\% disagreed with the statement. These responses were somewhat surprising given the limitations of GitHub Classroom (see Section~\ref{sec:limitations}) when compared to specialized AATs used by the students. Again, this difference may be due to the fact that the other AATs used by the students kept the teacher's tests hidden, whereas this approach allowed students to view, run, and debug the tests locally.

At the end of the survey, students were asked two open-ended questions regarding the main difficulties they encountered during the exercise and their suggestions for improving this type of exercise.

Students reported various difficulties when completing the exercise. Many struggled with understanding the observer-observable pattern and the correct implementation approach, as highlighted by one student: ``\textit{Understand how it was supposed to be implemented}''. Several students found it challenging to diagnose why certain tests failed, with one noting: ``\textit{It was difficult to understand most of error messages returned by the tests}'' and another stating ``\textit{I feel that when I make a mistake, the feedback that is given is insufficient}''. Technical issues were also common, particularly with the \texttt{Provider} library setup (responsible for DI), which caused the emulator to crash if not correctly configured. One student also pointed out the difficulty of dependency injection as a new concept. Additionally, there were concerns about the accuracy of provided test codes and difficulties running builds on GitHub. Despite these challenges, some students reported no major issues, indicating a diverse range of experiences.

Students provided several suggestions for improving the exercise. They emphasized the need for better-defined and clearer instructions, possibly organized by themes. Improved feedback from the tests was a common request, with some students suggesting more granular and detailed error descriptions. There was also a desire for a greater number of smaller, more focused tests rather than fewer, larger ones, as one student noted: ``\textit{Tests should be more granular. Instead of 4 big tests, it could be 20 small tests}''. Another suggestion was to avoid making the tests overly restrictive regarding how the application should be implemented, as this led to forced adjustments. Overall, while acknowledging the current model's benefits, students called for more information in the assignment descriptions and more explicit test requirements.

Complete survey results are available at \citep{alves_2024_13150846}.

\section{Discussion and conclusions}\label{sec_conclusion}

Regarding the research questions outlined in Section~\ref{sec_introduction}, the ``awesome quotes'' exercise described in Section~\ref{sec_experience} integrates Flutter, GitHub Classroom and GitHub Actions to enable the automatic assessment of a mobile programming assignment (\textbf{RQ1}). GitHub Classroom efficiently handles the student roster, sets up student repositories with appropriate permissions, and includes assignment instructions and initial code. The automatic assessment is enabled by leveraging the benefits of Flutter tests and GitHub Actions. Flutter allows integration tests to run without launching an emulator, significantly reducing the CPU cycles needed for execution on GitHub Actions, which has a time cap on execution. To the best of our knowledge, from the analyzed frameworks, only flutter\_test can run integration tests without an emulator, since the other solutions (Appium, Espresso, XCUITest, and Selendroid) rely heavily on the native UI framework of the respective platforms, requiring an actual device or emulator to perform interactions with the plaform's UI. As demonstrated in Section~\ref{sec:execution_quotas}, using an emulator for integration tests (as required by native approaches) would likely exceed the execution quota. GitHub Actions automate the build process, code quality validation, and test execution following a configuration file that can be easily adapted to different scenarios. However, we found GitHub Classroom's reporting capabilities to be limited, primarily due to inadequate feedback from standard GitHub Actions. This issue can be addressed by developing custom GitHub Actions to enhance auto-grading support.

The survey presented in Section~\ref{sec_results} answers \textbf{RQ2}, showing that these assessment model is well received by the students. It is worth noting that the majority of students prefer automated over manual assessment and consider this model suitable for both small exercises and projects, although they show a preference for the former. The ability to run tests locally is highly motivating, and we believe this is a key reason why this model is perceived as superior to specialized AATs, which conceal the tests from students. A possible implication of this is that giving access to the code might reduce the need for more detailed feedback, since students can use the test code to fully debug their code, similarly to what would happen in professional practice.

We have presented an assessment model that integrates GitHub Classroom, GitHub Actions, and the Flutter framework to automate the evaluation of mobile programming assignments. This model addresses the unique challenges of mobile development courses, including the complexity of development environments and the need for comprehensive testing of interactive, graphical applications. Future improvements could involve developing custom GitHub Actions to offer more detailed feedback and refined grading capabilities.
Overall, the proposed model demonstrates significant potential for improving mobile programming education by providing a more consistent and objective grading process, reducing the manual workload for instructors, and enabling students to receive timely and detailed feedback.

\section*{Abbreviations}\label{sec_abbreviations}

\begin{description}
    \item[AAT] Automated Assessment Tool
    \item[API] Application Programming Interface
    \item[CI] Continuous Integration
    \item[DI] Dependency Injection
    \item[IDE] Integrated Development Environment
    \item[LMS] Learning Management System
    \item[OS] Operating System
    \item[PR] Pull Request
    \item[SDK] Software Development Kit
    \item[UI] User Interface
\end{description}

\section*{Declarations}

\subsection*{Availability of data and materials}

Survey results are available at \href{https://doi.org/10.5281/zenodo.13150846}{https://doi.org/10.5281/zenodo.13150846}
GitHub repository of the assignment is available at \href{https://github.com/palves-ulht/awesome_quotes_exercise}{https://github.com/palves-ulht/awesome\_quotes\_exercise}

\bibliography{sn-bibliography}

\end{document}